\documentclass[
 reprint,
 amsmath,amssymb,
 aps,
showkeys,
]{revtex4-1}
\usepackage{enumerate}
\usepackage{siunitx}
\usepackage{graphicx}
\usepackage{dcolumn}
\usepackage{braket}
\usepackage{bm}
\usepackage{color}
\newcommand{\beginsupplement}{%
         \setcounter{table}{0}
         \renewcommand{\thetable}{S\arabic{table}}%
       \setcounter{figure}{0}
        \renewcommand{\thefigure}{S\arabic{figure}}%
    }

\newcolumntype{C}{ >{\centering\arraybackslash} m{2cm} }

\begin{document}


\title{Tracking the evolution from isolated dimers to many-body entanglement in NaLu$_x$Yb$_{1-x}$Se$_2$}
\author{Luke~Pritchard~Cairns$^1$, Ryan~Day$^1$, Shannon~Haley$^1$, Nikola~Maksimovic$^1$, Josue~Rodriguez$^1$, Hossein~Taghinejad$^1$, John~Singleton$^2$ and James~G. Analytis$^1$}

\affiliation{$^1$Department of Physics, University of California, Berkeley, California 94720, USA and Materials Science Division, Lawrence Berkeley National Laboratory, Berkeley, California 94720, USA \\
$^2$National High Magnetic Field Laboratory, Los Alamos National Laboratory,
Los Alamos, New Mexico 87545, USA}

\date{\today}

\begin{abstract}
We synthesise homogeneous compositions of NaLu$_x$Yb$_{1-x}$Se$_2$, connecting non-magnetic NaLuSe$_2$ to the triangular lattice spin liquid candidate NaYbSe$_2$. Thermal and magnetic properties are studied as the system evolves from one with dilute magnetic defects to one of a dense magnetic lattice. The field and temperature dependent heat capacity show the carriers of entropy crossover from isolated magnetic ions to a correlated lattice borne from spin dimers. For the dilute system we estimate the single ion anisotropy $(g_\perp/g_\parallel =3.13)$ and also the dimer exchange couplings $J_\parallel(=5.4$~K) and $J_\perp(=9.6$~K), in order to draw comparison to the half-doped and full magnetic compounds.

\end{abstract}

\pacs{Valid PACS appear here}
\maketitle

\section{\label{sec:Intro}Introduction}

The quantum spin liquid (QSL) is a theoretical construct wherein a system of strongly interacting localised moments shows no magnetic order but instead forms a massively entangled state, in some cases giving rise to fractionalised, dispersive excitations \cite{Savary2016,Zhou2017,Broholm2020}. A material's QSL candidacy often begins with a magnetic structure that conforms to some known model. For instance, the triangular lattice antiferromagnet (TLAF) plays host to a variety of QSL states \cite{Sachdev1992,Moessner2001,Zhu2015,Zhu2018,Li2020}, all of which rely on some form of anisotropic exchange or next nearest neighbour interactions \cite{Huse1988}. The candidate material should then exhibit no long range order (nor spin freezing) in spite of strong magnetic correlations, and finally, show some definitive, positive experimental signature of QSL physics. For example, a finite residual linear term in the heat capacity or thermal conductivity \cite{Yamashita2010,Yamashita2011}, a broad spectrum of low energy excitations observed via neutron scattering \cite{Han2012,Shen2016}, or a quantised thermal Hall conductivity \cite{Kasahara2018} have all previously been taken as evidence. Unfortunately however, at present, the most promising candidate materials remain subject to debate \cite{Ni2019,Zhu2017_mimicry,Kimchi2018,Han2016,Czajka2021}. 

NaYbSe$_2$ is one such candidate QSL material. It belongs to a recently discovered class of Yb delafossites \cite{Liu2018,Baenitz2018}, and comprises alternating planes of distorted YbSe$_6$ and NaSe$_6$ octahedra (as shown in the inset of Figure~\ref{Homogeneity}~(a)). A combination of spin-orbit coupling and crystalline electric field (CEF) gradients split the Yb$^{3+}$ 4f$^{13}$ energy levels into seven Kramers doublets, the lowest of which is separated from the first excited state by $\sim180$~K \cite{ZhangCEF2021}. At low temperatures the Yb$^{3+}$ ions are therefore confined to the doubly degenerate ground state, and can be modelled as pseudospin-1/2 moments \cite{Schmidt2021}. These magnetic sites make up a perfectly triangular lattice, and are coupled in an anisotropic, antiferromagnetic fashion - conforming to those anisotropic TLAF models which predict QSL states in certain pockets of parameter space \cite{Zhu2018}. Indeed, previous studies have observed a finite Sommerfeld coefficient in the specific heat \cite{Ranjith2019}, low-energy excitations with a V-shaped dispersion about the $\Gamma$-point \cite{Dai2021}, and persistent spin fluctuations down to 50~mK \cite{Zhang2021_muSR}. Altogether this constitutes a strong case for a ground state with gapless, dispersive excitations. However - as with the majority of QSL candidates - these excitations do not appear in the low-temperature thermal conductivity \cite{Zhu2021}.


We propose a divergence from the typical exploration of a QSL candidate material. By first stripping away the magnetic interactions it is possible to isolate the single ion behaviour within an identical CEF environment. Then, by steadily scaling these back up again we have the power to investigate the evolution of correlations, first in single dimers, through larger finite clusters, and building towards the full magnetic state. 


In this work we have succeeded in growing NaYbSe$_2$, its non-magnetic analogue NaLuSe$_2$, and also  doped NaLu$_x$Yb$_{1-x}$Se$_2$, with $x=0.5$ and 0.9. By employing a range of experimental techniques we are able to demonstrate that the doped compounds are high quality single crystals, both of which retain the salient features of the parent compound and - crucially for any meaningful study into the evolution of correlations  - comprise a homogeneous mixture of magnetic and non-magnetic ions. The four compounds were chosen to act as, respectively, the QSL candidate material $(x=0)$, the non-magnetic background $(x=1)$, a system at the percolation threshold $(x=0.5)$, and a system of isolated magnetic ions and small clusters $(x=0.9)$. From this, we are able to investigate the evolution of entanglement as a function of magnetic cluster size via measurements of the low-temperature specific heat. We believe that NaLu$_x$Yb$_{1-x}$Se$_2$ offers a new route to an understanding of similar candidate materials, in which spin-orbit coupling is leveraged as a means to generate anisotropic exchange interactions, but the resultant hugely composite, pseudospin moments may not simply map onto the archetypal QSL models.


\section{\label{sec:Results}Results}

Detailed descriptions of the crystal growth and experimental methods are included in the supplementary material (SM). 


\subsection{\label{subsec:Samp}Sample Characterisation and Evidence for Homogeneity}

Single crystals of NaLu$_x$Yb$_{1-x}$Se$_2$ (with $x=0$, 0.5, 0.9 and 1)  were grown by the flux method, following a similar procedure to previous studies \cite{Ranjith2019}. Energy dispersive X-ray (EDX) spectroscopy was employed to confirm that all four compounds are of the desired composition, whilst powder X-ray diffraction (PXRD) shows that all share the anticipated R$\overline{3}$m rhombohedral structure. Further, as shown in Figure~\ref{PXRD}, with increasing Lu content the in-plane lattice constant decreases, whilst the out-of-plane increases. This is typical of similar compounds containing Yb/Lu in an octahedral environment \cite{Li2015,Liu2018}. We note however that both techniques are essentially averaging over too large a length scale to definitively claim microscopic homogeneity, although a lack of broadening in the diffraction peaks might be taken as weak evidence. 

\begin{figure}[h]
\centering
\includegraphics[width=0.45\textwidth]{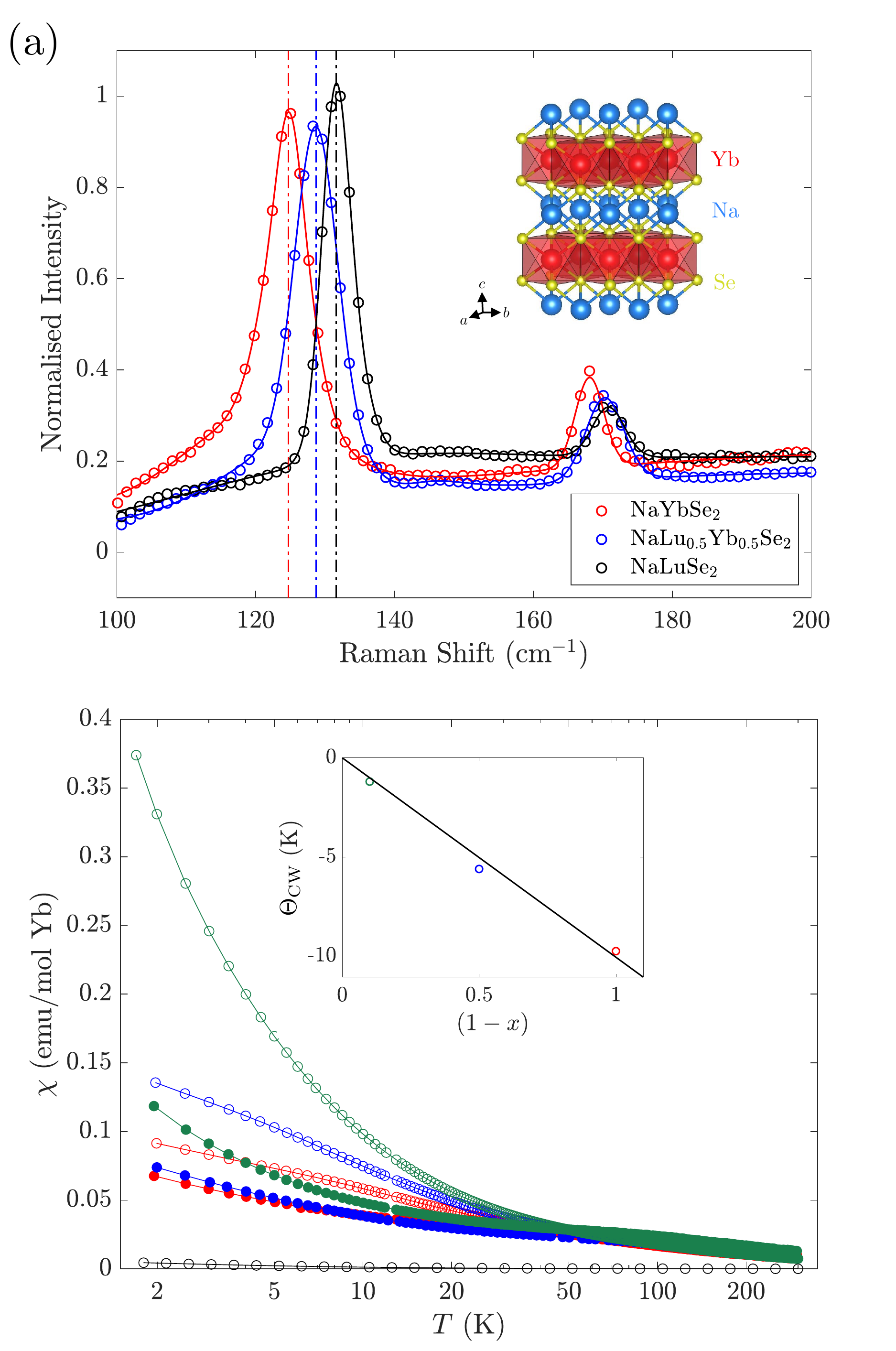}
\caption{(a) Raman spectra of the materials NaLu$_x$Yb$_{1-x}$Se$_2$, where $x=0$, 0.5 and 1, obtained at 300~K in all cases. The solid lines show a sum of Gaussians fit through the data, while the dot-dash lines show the position of the $E_g$ peak. The inset shows the crystal structure of NaYbSe$_2$ (R$\overline{3}$m, space group 166). (b) Magnetic susceptibility per mole of Yb for NaLu$_{x}$Yb$_{1-x}$Se$_2$, where ${x=0}$, 0.5, 0.9, and per mole of material for NaLuSe$_2$, plotted as a function of temperature for both in-plane and out-of-plane fields. All data was obtained in a field of 0.1~T. The inset shows the linear evolution of the Curie-Weiss temperature as a function of Yb content (estimated from fits in the region $T<50$~K, shown in Figure~\ref{Susceptibility_SM}~(a)).}
\label{Homogeneity}
\end{figure}

Instead, Raman spectroscopy - which probes the $E_g$ and $A_{1g}$ phonon modes \cite{ZhangCEF2021} - can effectively demonstrate homogeneity. As shown in Figure~\ref{Homogeneity}~(a), the peak energy of the $E_g$ mode is $\sim$7~cm$^{-1}$ (10~K) larger in NaLuSe$_2$ as compared to NaYbSe$_2$ - this implies a stiffening of the bonds in the non-magnetic compound. If the half-doped compound comprised separate regions of the two end compounds, we would expect two distinct $E_g$ peaks in the Raman spectrum. Instead, we observe a single sharp peak lying almost exactly between those of the two end compounds. Considering the short-range nature of the Raman vibrations, our measurements reflect the uniform intermixing of magnetic ions without a detectable elemental segregation. We note that previous studies have used Raman spectroscopy for a similar purpose \cite{Hernandez2005,Su2014}.

Further evidence for microscopic homogeneity comes from the magnetisation data. Figure~\ref{Homogeneity}~(b) shows the magnetic susceptibility plotted as a function of temperature. In all cases, the zero-field and field cooled curves are identical. Regardless of doping, the magnetic compounds exhibit similar features for an in-plane field. At higher temperatures ($T>100$~K), all are well described by the Curie-Weiss law, with a Curie-Weiss temperature ($\Theta_{\text{CW}})$ $\sim50$~K and a moment of $\sim4.5\mu_B$ (as shown in Figure~\ref{Susceptibility_SM}~(a)). This is similar to the value for a Yb$^{3+}$ free ion $(=4.54\mu_B$). In all cases, the trigonal CEF splits the Yb$^{3+}$ $j=7/2$ energy level into 4 Kramers doublets, with the first excited state being approximately $180$~K above the ground state \cite{ZhangCEF2021}. This leads to the emergence of a pseudospin-1/2 state at the lowest temperatures, wherein the Yb$^{3+}$ ion is constrained to the ground state doublet. Strong second order corrections complicate the analysis in this temperature regime (as discussed in the SM), but the in-plane inverse susceptibility is well fit using a modified form of the Curie-Weiss law for $T<50$~K (Figure~\ref{Susceptibility_SM}). Crucially, the extracted low-temperature $\Theta_{\text{CW}}$ scales linearly with Yb content (as shown in the inset of Figure~\ref{Homogeneity}~(b)), suggesting that the strength of the mean field at the Yb sites - and therefore the average number of magnetic nearest neighbours - scales similarly. This is the expectation for a homogeneous lattice \cite{Spalek1986}. Again, if there were instead separate macroscopic regions of the two end compounds, $\Theta_{\text{CW}}$ would remain constant.

\subsection{\label{subsec:Lu9Yb1}NaLu$_{0.9}$Yb$_{0.1}$Se$_2$}

\begin{figure}[h!]
\centering
\includegraphics[width=0.4\textwidth]{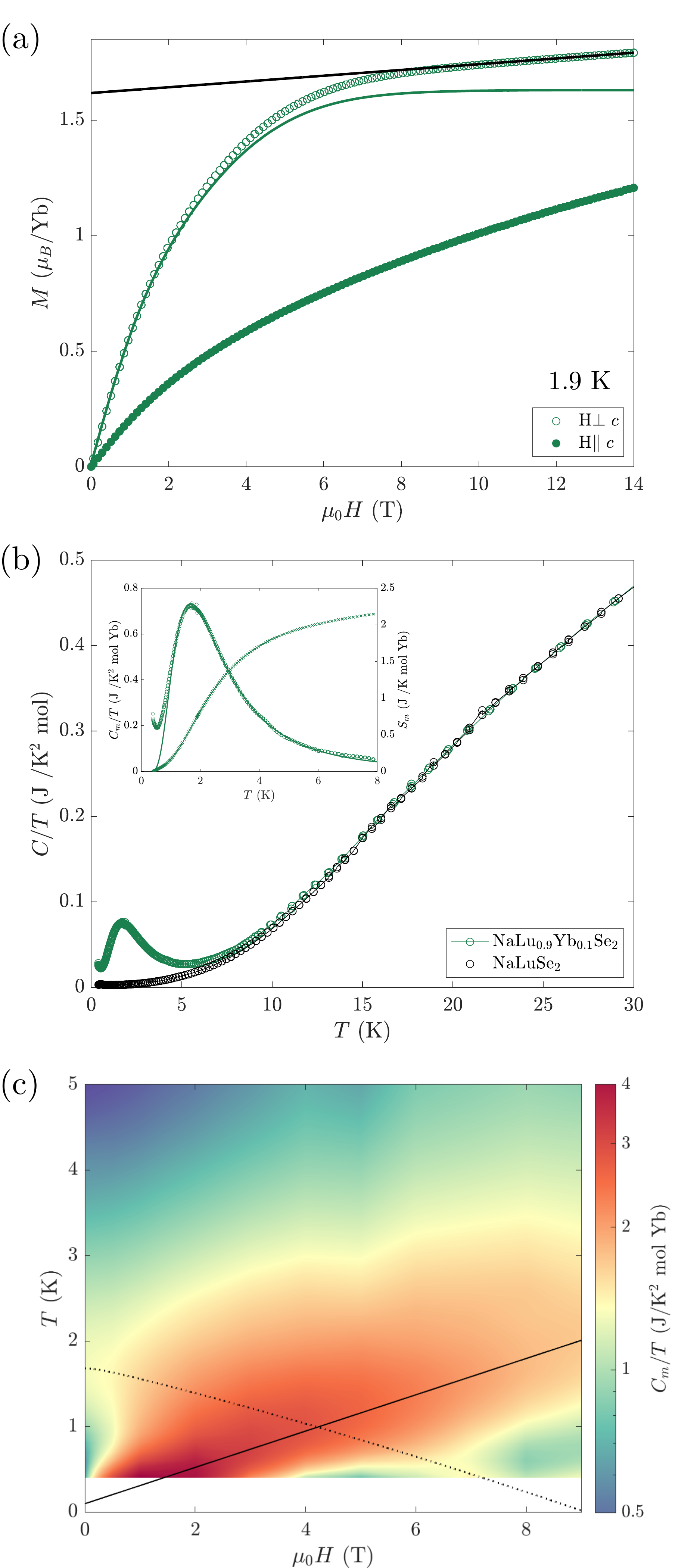}
\caption{(a) Magnetisation as a function of field for NaLu$_{0.9}$Yb$_{0.1}$Se$_2$, plotted for both field directions. Both data sets were taken at 1.9~K. The black line shows the linear Van Vleck contribution for in-plane fields, estimated by fitting to the $H>12$~T data. The green line shows the in-plane magnetisation with this contribution subtracted. (b)~Zero-field heat capacity per mole of material for NaLu$_{0.9}$Yb$_{0.1}$Se$_2$ and NaLuSe$_2$. The inset shows the magnetic heat capacity of the former, and also the released entropy. The solid line is a fit to a two-level system, with $\Delta E=5.4$~K and 37\% of the sites contributing. (c)~Specific heat as a function of temperature and field for NaLu$_{0.9}$Yb$_{0.1}$Se$_2$. The applied field was out-of-plane. The solid/dotted lines show the approximate positions of the lowest energy single ion and isolated dimer peak respectively. The colour bar is on a logarithmic scale for clarity.}
\label{Lu9Yb1}
\end{figure}

As shown in Figure~\ref{Lu9Yb1}~(a), the the magnetisation of NaLu$_{0.9}$Yb$_{0.1}$Se$_2$ is highly anisotropic and fails to saturate for either field direction. For in-plane fields, the high-field magnetisation is linear (as clearly demonstrated by differentiating the data with respect to field, Figure~\ref{Magnetisation_SM}~(a)). This is the linear Van Vleck paramagnetism - related to field-induced transitions between the CEF split levels \cite{Magnetochemistry} - which we estimate to be 0.0125~$\mu_B\text{T}^{-1}/$Yb. From this, we can extract the saturation moment ${M^s_{\perp}=1.63\mu_B}/$Yb, and therefore ${g_{\perp}=2\frac{M_{\perp}^s}{\mu_B}=3.26}$. This is approximately 10\% larger than the in-plane $g$-factor determined previously for the full magnetic compound \cite{Ranjith2019,ZhangModel2021,ZhangCEF2021}. A similar analysis is not possible for out-of-plane fields, as the magnetisation remains non-linear up to 14~T, and the {Van Vleck} contribution is not expected to be simply a small additive correction \cite{Pocs2021}. Instead, $g_{\parallel}$ will be estimated from a field induced Schottky anomaly in the specific heat, as described below. 

Regardless, a strong anisotropy is clearly evident. The magnetisation in NaLu$_{0.9}$Yb$_{0.1}$Se$_2$ will be dominated by isolated Yb$^{3+}$ ions, and the anisotropy is thus primarily a single ion phenomenon. This supports the qualitative picture, wherein the prolate 4f$^{13}$ quadrupolar moment is sandwiched by negative charge and therefore preferentially orients in the plane \cite{Rinehart2011,Zangeneh2019}. 



Figure~\ref{Lu9Yb1}~(b) shows the zero-field specific heat of NaLu$_{0.9}$Yb$_{0.1}$Se$_2$, plotted alongside the non-magnetic NaLuSe$_2$. The latter behaves as an almost ideal non-magnetic analogue, with a featureless heat capacity that roughly follows the Debye law ($\Theta_D=259.8$~K). At the lowest temperatures there appears a small increase (see Figure~\ref{LuHC}~(b)), which we may attribute to the interaction between the Lu nuclear quadrupolar moment and the surrounding CEF gradients. Above approximately 15~K, the two compounds fall onto the same curve. We expect some excess heat capacity in the magnetic compound at higher temperatures due to thermal population of the Yb$^{3+}$ first excited CEF level \cite{ZhangModel2021}, but that does not affect the low-temperature behaviour. 

Below 10~K a peak feature develops in NaLu$_{0.9}$Yb$_{0.1}$Se$_2$. This is shown more clearly in the inset, which plots the magnetic specific heat per mole of Yb, obtained by subtracting an interpolation through the NaLuSe$_2$ data. Also included in the inset is the released entropy, which rises to 2.15~JK$^{-1}$mol$^{-1}$ at 8~K. This is only $\sim$37\% of the $R\ln(2)$ entropy contained within a system of spin-1/2 moments. Instructively, this is also approximately the same percentage of sites that will have one magnetic nearest neighbour in a triangular lattice at 10\% filling. The observed peak is therefore likely attributable to the entropy stored in Yb dimers. As a first approximation, the peak is well fit by a simple two-level model (shown in the figure) - implying that these dimers interact via a well defined energy scale.

In order to investigate the missing entropy, Figure~\ref{Lu9Yb1}~(c) shows the temperature-field evolution of the specific heat (the same data is also presented in Figure~\ref{HC_SM}~(a) as a line plot). With the application of field, there appears a second peak superimposed on the first, which moves to higher temperatures with increasing field. Significantly, the released entropy rises almost to the anticipated $R\ln(2)$ at intermediate fields. This implies that the remainder of the (non-dimerised) magnetic ions contribute this second peak, and therefore hold the missing entropy.

The field induced peak is naturally explained by a Zeeman splitting of the Yb$^{3+}$ ground state doublet, for which the energy splitting ${\Delta=g_{\parallel}\mu_BH_{\text{eff}}}$. $H_{\text{eff}}=\sqrt{H^2+H_0^2}$, where $H$ is the applied field and $H_0$ accounts for a dipolar field from neighbouring ions, or possibly impurity spins \cite{Bag2021,Rau2018}. This additional contribution is necessary to explain the zero-field splitting, which is evident from a low-temperature upturn in the zero-field data. We note that any second order corrections to the Zeeman splitting will affect both energy levels in a similar fashion, such that the energy splitting remains linear in field. 

The picture is therefore that of two non-interacting species, one of single ions and the other of dimers, each of which have an independent characteristic energy scale. In order to draw quantitative conclusions, we can attempt to model the specific heat within this framework. For the most general discretised system - having $n$ levels, each with an energy $E_n$ and degeneracy $g_n$ - the specific heat is given by
\begin{equation}
C(T) = R \frac{d}{dT} \bigg[\frac{\sum_n g_n E_n e^{-E_n/k_B T}}{\sum_n  g_n e^{-E_n/k_B T}}\bigg],
\label{C1}
\end{equation}
which in the case of a Zeeman split doublet reduces to
\begin{equation}
C(T)=R\bigg(\frac{\Delta}{T}\bigg)^2 \frac{ e^{-\Delta/T}}{(1+e^{-\Delta/T})^2},
\label{C2}
\end{equation}
where the energy splitting $\Delta$ is given above (and expressed here in Kelvin). This function has a maximum at a temperature of ${~0.42\Delta}$, and the field evolution of the specific heat can therefore be used to determine the out-of-plane $g$-factor. Figure~\ref{Lu9Yb1}~(c) shows a linear fit through the peak, from which we estimate $g_{\parallel}=1.04$ (the same linear fit is shown more clearly in the inset of Figure~\ref{HC_SM}~(a)). Together with $g_\perp=3.26$ estimated from the magnetisation data, we calculate an easy-plane anisotropy of 3.13. This is similar to (but slightly larger than) the easy-plane anisotropy estimated for the full magnetic compound \cite{Ranjith2019,ZhangCEF2021,ZhangModel2021}.

In order to model the remainder of the magnetic ions, we treat them as a system of isolated dimers. For a single dimer, the four possible states can be expressed $\ket{S,s_z}$, where $S$ is the total effective spin and $s_z$ is the projection of this spin along the quantisation axis. For an isotropic antiferromagnetic exchange interaction, the ground state singlet ($\ket{0,0}$) will be separated from the degenerate triplet ($\ket{1,-1}$, $\ket{1,0}$ and $\ket{1,1}$) by an energy proportional to the strength of exchange. However, in the case of anisotropic exchange, the triplet level will in general be split. For example, consider anisotropic exchange described by the Hamiltonian
\begin{equation}
    \mathcal{H}=J_{\perp} (S^x_1 S^x_2 + S^y_1 S^y_2) + J_{\parallel} S^z_1 S^z_2,
\end{equation}
or the XXZ model. In this scenario, the $\ket{1,0}$ state will be separated from the ground state singlet by an energy $J_{\perp}$, whereas the $\ket{1,\pm1}$ doublet will be separated by $J_{\parallel}$. Crucially, an applied field $H$ will act to shift the energies by $-gs_z\mu_BH$, implying that the $\ket{1,\pm1}$ doublet will be split, but the other energy levels will stay constant. In principle therefore, the field evolution of the specific heat can be used to determine the strength, sign and anisotropy of the dimer exchange interaction. 

We now apply this analysis to NaLu$_{0.9}$Yb$_{0.1}$Se$_2$. We model the isolated dimers as above, and also include an independent population of isolated single ions. The fitting function is discussed in the SM, but derives from Equation~\ref{C1}, with the various energy levels defined by $g_\parallel$ (which we assume to be similar for both species), $J_{\perp}$, $J_{\parallel}$ and also the applied field. Higher order corrections to the Zeeman splitting are somewhat relevant here, and lead to non-linear, asymmetric splitting of the triplet $\ket{1,\pm1}$ level. 

\begin{figure}[h!]
\centering
\includegraphics[width=0.4\textwidth]{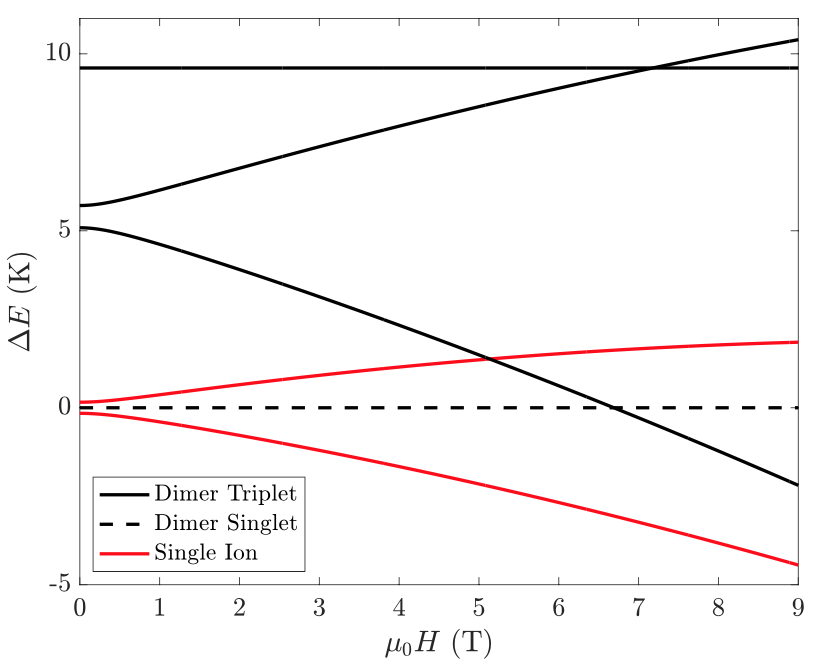}
\caption{Field evolution of the energy levels for both a isolated single ion (red) and isolated dimer (black) calculated from the parameters estimated by the magnetisation ($g_\parallel=1.04$), in-field specific heat ($J_\parallel=5.4$~K and $J_\perp=9.6$~K) and susceptibility ($\alpha_\perp = 0.016~\mu_B$T$^{-1}$, see SM).}
\label{Elevels}
\end{figure}

If we estimate the relative populations from the released entropy (at $\sim$37\% and $\sim$59\% respectively), this leaves only $J_{\perp}$ and $J_{\parallel}$ to be determined. However, certain aspects of the data can help to put constraints on these. For example, the position of the peak in the zero-field specific heat implies that the lowest energy splitting of the dimer subsystem is approximately 5~K without an applied field. Further, the observed low-temperature upturn at higher fields (and corresponding decrease of released entropy within the measured temperature window - Figure~\ref{HC_SM}~(b)) implies there should be a small energy splitting at high fields. This can only arise due to a crossing of the Zeeman shifted $\ket{1,1}$ state and the singlet ground state (as shown more clearly in Figure~\ref{Elevels}). Given $g_\parallel =1.04$, the $\ket{1,1}$ level must be at $\sim5$~K in zero-field in order to give the observed upturn at higher fields. This naturally accounts for the zero-field peak, and implies that $\ket{1,\pm1}$ is the first excited state.


Fitting to the zero-field data gives $J_{\parallel}=5.4$~K and $J_{\perp}=9.6$~K, although the latter is significantly less well constrained as the smaller energy splittings dominate the specific heat. The field evolution of the isolated ion and dimer energy levels is shown Figure~\ref{Elevels}. This model is able to reproduce the essential features of the in-field specific heat data, as shown in Figure~\ref{HC_SM}~(b).  The magnitudes of the exchange couplings are slightly larger than those estimated for the full magnetic compound (via fits to the low-temperature susceptibility \cite{Ranjith2019,ZhangModel2021}), but of the same order and proportionality, and with the same hierarchy.

We note that there may be some additional complications to the above model. For example, this framework does not account for larger magnetic clusters, although the released entropy alongside the adequacy of the zero-field fitting would suggest that these are insignificant. Also, there may be relevant anisotropic exchange terms beyond the XXZ model, but previous studies disagree on whether these should be negligible in comparison \cite{ZhangModel2021,Schmidt2021}. Finally, the field evolution will necessarily be impacted by second order corrections to the Zeeman splitting. The strength of these corrections have been determined by fitting to the susceptibility (Figure~\ref{Susceptibility_SM}), but this appears to give a slight underestimation as compared to the same correction extracted from the magnetisation (although it is only possible to draw comparison in one case).

\subsection{\label{subsec:Lu5Yb5}NaLu$_{0.5}$Yb$_{0.5}$Se$_2$ and NaYbSe$_2$}

\begin{figure}[h!]
\centering
\includegraphics[width=0.4\textwidth]{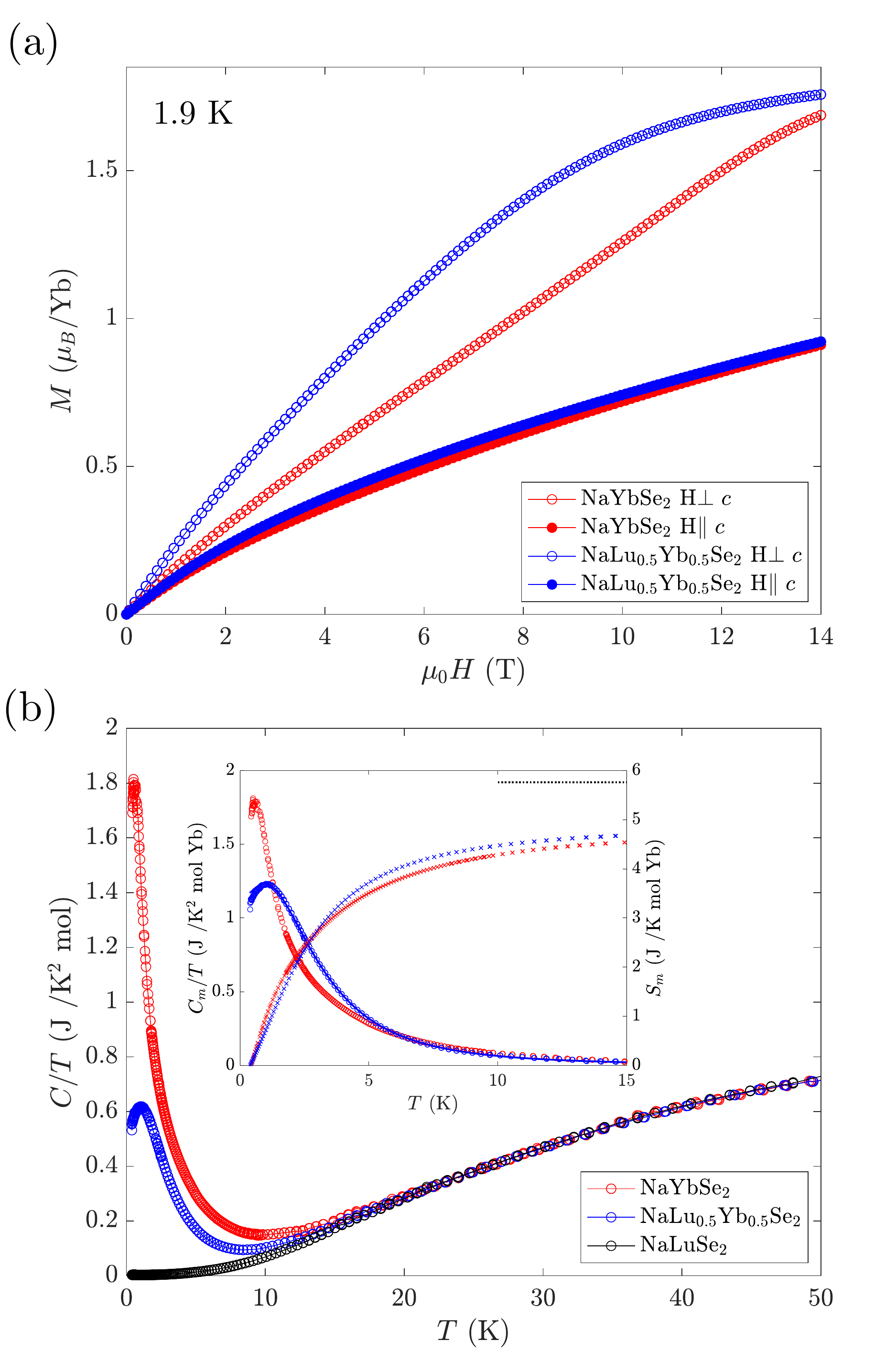}
\caption{(a) Magnetisation of NaLu$_{0.5}$Yb$_{0.5}$Se$_2$ and NaYbSe$_2$ as a function of field. (b) Specific heat of both compounds, compared to the non-magnetic analogue NaLuSe$_2$. The inset shows the magnetic specific heat and also the released entropy. The solid blue line is a fit of the NaLu$_{0.5}$Yb$_{0.5}$Se$_2$ data to a Gaussian broadened two-level system, and the dashed black line is at $R\ln{2}$, the upper bound for a spin-1/2 system. }
\label{Lu5Yb5}
\end{figure}

The data for NaLu$_{0.5}$Yb$_{0.5}$Se$_2$ and NaYbSe$_2$ are largely similar, and it seems sensible to discuss them in tandem. Figure~\ref{Lu5Yb5}~(a) shows the magnetisation. Both compounds retain a strong anisotropy, and neither approach linearity at high-fields. In fact, as demonstrated in Figure~\ref{Magnetisation_SM}~(b), the out-of-plane magnetisation in NaLu$_{0.5}$Yb$_{0.5}$Se$_2$ fails to saturate even for fields as large as 40~T. Similar to NaLu$_{0.9}$Yb$_{0.1}$Se$_2$, this would imply that the Van Vleck contribution is not simply a negligible addition, and must be determined via a different means (for example by fitting to the susceptibility, as shown in Figure~\ref{Susceptibility_SM}). Without saturation it is difficult to be quantitative, but it is perhaps interesting that the low-field anisotropy falls with increasing Yb content, suggesting that the single ion anisotropy is potentially mitigated.

As with NaLu$_{0.9}$Yb$_{0.1}$Se$_2$, the zero-field specific heat (shown in Figure~\ref{Lu5Yb5}~(b)) of both compounds exhibits a low-temperature peak, before falling onto the NaLuSe$_2$ curve above 20~K. In this case however, the magnetic specific heat cannot be fit using a simple two-level model. Instead, a `best' fit to the NaLu$_{0.5}$Yb$_{0.5}$Se$_2$ data (shown in the inset of Figure~\ref{Lu5Yb5}~(b)) requires a significant Gaussian broadenening of the two-level system, although with a similar energy splitting ($\sim5$~K) to the sharp two-level description of NaLu$_{0.9}$Yb$_{0.1}$Se$_2$. This implies that the isolated dimer model is no longer relevant, and may reflect a smearing of the relevant energy scale. The NaYbSe$_2$ peak cannot be interpreted within any similar framework, perhaps suggesting that the idea of discretised energy levels is no longer applicable. In both cases however, almost the full anticipated $R\ln{2}$ entropy is released over the measured temperature window (also shown in the inset of Figure~\ref{Lu5Yb5}~(b)), and the majority of magnetic sites must therefore contribute to the low-temperature peak.

The number of isolated ions can once again be estimated from the in-field specific heat data (Figure~\ref{HC_SM}). For NaLu$_{0.5}$Yb$_{0.5}$Se$_2$, there appears a field induced Schottky anomaly at low temperatures. However, this feature releases only an additional $0.05R\ln{2}$ of entropy, corresponding to 5\% of the magnetic ions being essentially uncorrelated. A simple calculation would suggest that only 1.5\% should be without a magnetic nearest neighbour, but the number of effective single ions might be increased due to disorder (induced via Na substitution at the magnetic sites or excess Se, as implied by EDX measurements). Similar to the zero-field data, the field evolution can be roughly fit by considering that $\sim$90\% of the sites participate in a significantly Gaussian broadened two-level system, whilst $\sim$5\% remain as isolated single ions, and are therefore Zeeman split. The remaining 5\% can be accounted for again due to the excess Na and Se, which might result in a slight overestimation of the number of moles of magnetic ions.

The in-field data for NaYbSe$_2$ is likewise intractable (Figure~\ref{HC_SM}~(d)), but appears to evolve in a similar manner, with an increasing energy scale and broadening of the peak. Any quantitative analysis is further confused by a third peak at higher fields, which has been interpreted previously as the onset of antiferromagnetic order \cite{Ranjith2019}. Importantly however, the small size of the field induced Schottky anomaly in NaLu$_{0.5}$Yb$_{0.5}$Se$_2$ (and the absence of one in NaYbSe$_2$) would imply that in both compounds the overwhelming majority of sites are not interacting with the field in the way that a single ion might. Further, the broadening of the energy scale in NaLu$_{0.5}$Yb$_{0.5}$Se$_2$, and the absence of a significant low-temperature upturn at higher fields in either (as observed in NaLu$_{0.9}$Yb$_{0.1}$Se$_2$), would imply that neither compound can be described as a system of isolated dimers.

\section{\label{sec:Discussion}Discussion}

An anomalous peak in the zero-field specific heat is a universal feature of frustrated systems that exhibit no magnetic order. Similar features have been observed to occur regardless of the nature of the spinful object \cite{Nakatsuji_2007,Balz2017,Rawl2017,Ramirez1991}, magnetic structure \cite{Ma2020} or even dimensionality \cite{Okamoto2007}. The anomalous peak in NaYbSe$_2$ has variously been interpreted as arising due to strong spin fluctuations \cite{ZhangModel2021}, correlations between quasi-static spins \cite{Zhu2021} or simply as a consequence of the TLAF fully frustrated, disordered ground state \cite{Ranjith2019}. In fact, almost any disordered ground state will result in a specific heat which rises from zero-temperature and falls with a $1/T^2$ dependence at higher temperatures, such that a peak is a fairly generic feature, and the vast majority of interpretations remain valid. The peak position will invariably give some measure of the strength of correlations, but in order to obtain real insight it is often necessary to track the evolution of the specific heat under some perturbation. For example, the application of field should immediately reveal any uncorrelated spins via the emergence of an additional Schottky peak. The fact that this is absent in NaYbSe$_2$ already implies that all of the spins are participating in some form of correlated state. 


An alternative method of perturbing the system is through doping \cite{Ramirez1992}, which in principle allows a tuning of the magnetic correlations. Whilst doping might introduce appreciable disorder, we note that this is not necessarily an anathema to the QSL state \cite{Ma2020,Furukawa2015}. Upon doping NaYbSe$_2$, the anomalous zero-field peak evolves into a well characterised two-level Schottky peak in the dilute system (NaLu$_{0.9}$Yb$_{0.1}$Se$_2$). We have demonstrated that this feature can be entirely attributed to isolated dimers, and therefore defines the relevant energy scale for singlet formation. Nearest neighbour interactions should not change significantly with doping, and indeed, this energy scale remains relevant in the half-doped compound. In this case however the feature is broadened, perhaps signaling the emergence of a random singlet phase. This phase can be induced by disorder - or pertinently, site dilution \cite{Singh2010} - and is characterised by a combination of orphan spins, isolated singlet-dimers and also clusters of resonating singlet-dimers, each of which interact via a different energy scale \cite{Kawamura2019}. Further, the random singlet phase has been predicted to mimic the signatures of some QSL states \cite{Kimchi2018}, despite it being a simple product state. For example, the low-energy excitations - such as singlet pair breaking or diffusive orphan spins - can give a $T$-linear low-temperature specific heat, or even a $T$-linear thermal conductivity. An observation of the latter in the half-doped compound would certainly be revealing, as NaYbSe$_2$ exhibits no such behaviour \cite{Zhu2021}.

Finally, the slightly reduced temperature of the peak in the half-doped compound as compared to NaLu$_{0.9}$Yb$_{0.1}$Se$_2$ might imply that larger sized clusters interact on a reduced energy scale. This would naturally explain why the peak in the full magnetic compound occurs at a lower temperature still. However, it is clear that the absence of a singlet peak (sharp or broadened) in NaYbSe$_2$ suggests that the interactions have evolved beyond two-site correlations, and perhaps even that we can preclude the random singlet phase.

\begin{figure}[b!]
\centering
\includegraphics[width=0.47\textwidth]{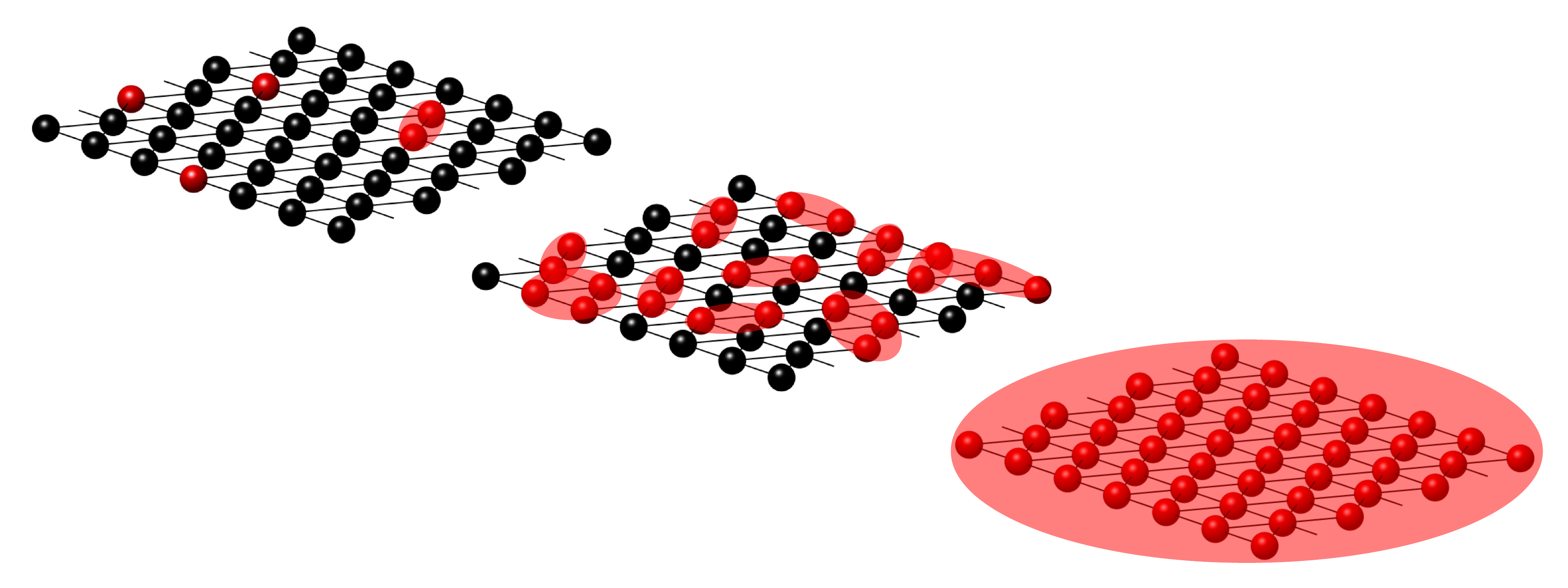}
\caption{Schematic to show the proposed evolution of correlations as a function of doping in NaLu$_{x}$Yb$_{1-x}$Se$_2$, from isolated single ions and dimers ($x=0.9$, left), though larger, finite magnetic clusters at the percolation threshold ($x=0.5$, middle), to the full magnetic, many-body entangled state ($x=0$, right).}
\label{Evolution}
\end{figure}

From the analysis of NaLu$_{0.9}$Yb$_{0.1}$Se$_2$, we are in a position to be more quantitative. Firstly, the easy-plane anisotropy is found to be slightly enhanced as compared to the full magnetic compound (3.13 compared to $\sim3$ \cite{Ranjith2019,ZhangCEF2021,ZhangModel2021}). This implies that the anisotropy observed in NaYbSe$_2$ is primarily a single ion phenomenon, but also that it is suppressed by increasing the average number of Yb nearest neighbours. The single ion anisotropy is driven by the surrounding CEF gradients \cite{Zangeneh2019}, and it seems likely therefore that this suppression is due to the different charge distributions of Lu$^{3+}$ and Yb$^{3+}$. In an octahedrally coordinated environment, Yb$^{3+}$ is expected to have a slightly larger ionic radius \cite{Shannon1976}. This would act to force the prolate 4f$^{13}$ quadrupolar moment out of the plane, and hence suppress the easy-plane anisotropy \cite{Rinehart2011}. 
\\

From the field evolution of the specific heat we have also estimated the pseudospin-1/2 dimer exchange constants $J_\parallel(=5.4$~K) and $J_\perp(=9.6$~K). These are slightly larger than those estimated for NaYbSe$_2$ \cite{Ranjith2019,ZhangModel2021}. This may be due to the frustration present in the full magnetic compound, or perhaps the simplifying assumptions of our model. However, unlike the magnetic susceptibility, the specific heat more directly reflects the energy spectrum, and should therefore be an accurate tool in determining the various energy scales. In refining this technique, it may be more fruitful to investigate lower dopings - wherein the impact of larger clusters can safely be disregarded - or in-plane fields - for which the second order corrections to the Zeeman splitting are smaller, and the data might therefore be more amenable to analysis. 

Finally, and perhaps ominously, we note that the singlet ground state of an isolated dimer is non-magnetic. We should therefore expect a reduction of the magnetic susceptibility below a temperature of the order of the exchange coupling. However we observe no divergence from paramagnetic behaviour, even when the susceptibility is measured at a significantly smaller field (Figure~\ref{Susceptibility_SM}~(c)). This may be due to single ions dominating the susceptibility - although a $\sim35$\% decrease should still be observable - or perhaps due to orphan spins induced by disorder \cite{Schiffer1997}, which can act to enhance the susceptibility at low temperatures \cite{Ramirez2000}. One alternative is that the singlet-triplet model is not sufficient to model a strongly spin-orbit coupled, pseudospin-1/2 dimer, but a similar analysis has been successful in previous studies of Yb dimer systems \cite{Nikitan2020}.


We have demonstrated that NaLu$_{x}$Yb$_{1-x}$Se$_2$ represents the ideal framework for understanding a recently discovered class of candidate spin liquid materials. The compounds are straightforward to grow, share similar characteristics across the full doping range, have an ideal non-magnetic analogue, and comprise a homogeneous mixture of magnetic and non-magnetic ions. This allows us to track the evolution of spin correlations from isolated dimers through small entangled clusters up to the many-body entangled state in the full magnetic compound. Such an evolution opens up the possibility of studying percolation transitions in QSL candidate materials.


Recently we became aware of \cite{Hausler2022}, which reports the synthesis and characterisation of a related compound, NaYb$_{1-x}$Lu$_{x}$S$_2$. We would like to thank {Ehud~Altman} and {Chunxiao~Liu} for some enlightening discussions. 

This work was supported by the U.S. Department of Energy, Office of Science, Basic Energy Sciences, Materials Sciences and Engineering Division under contract DE-AC02-05-CH11231 within the Quantum Materials program (KC2202). Work performed by N.M. was supported by the U.S. Department of Energy, Office of Science, Office of Basic Energy Sciences, Materials Sciences and Engineering Division under Contract No. DE-AC02-05-CH11231, within the Quantum Systems Accelerator Program. R.D. is currently supported by the Canadian Government under a Banting Fellowship. S.H. was supported as part of the Center for Novel Pathways to Quantum Coherence in Materials, an Energy Frontier Research Center funded by the United States Department of Energy, Office of Science, Basic Energy Sciences. J.R. was supported by the National Science Foundation through award no. DMR-1905397. H.T. was supported by the EPiQS Initiative of the Gordon and Betty Moore Foundation through grant no. GBMF9067. A portion of this work was performed at the National High Magnetic Field Laboratory (NHMFL), which is supported by National Science Foundation Cooperative Agreement No. DMR-1644779, the State of Florida and the Department of Energy (DOE). J. S. acknowledges support from the DOE BES program ``Science of 100~T".

%

\onecolumngrid
\newpage
\beginsupplement
\pagebreak

\section{Supplementary Material}

\subsection{\label{sec:Methods}Methods}

\begin{figure}[h]
\centering
\includegraphics[width=1\textwidth]{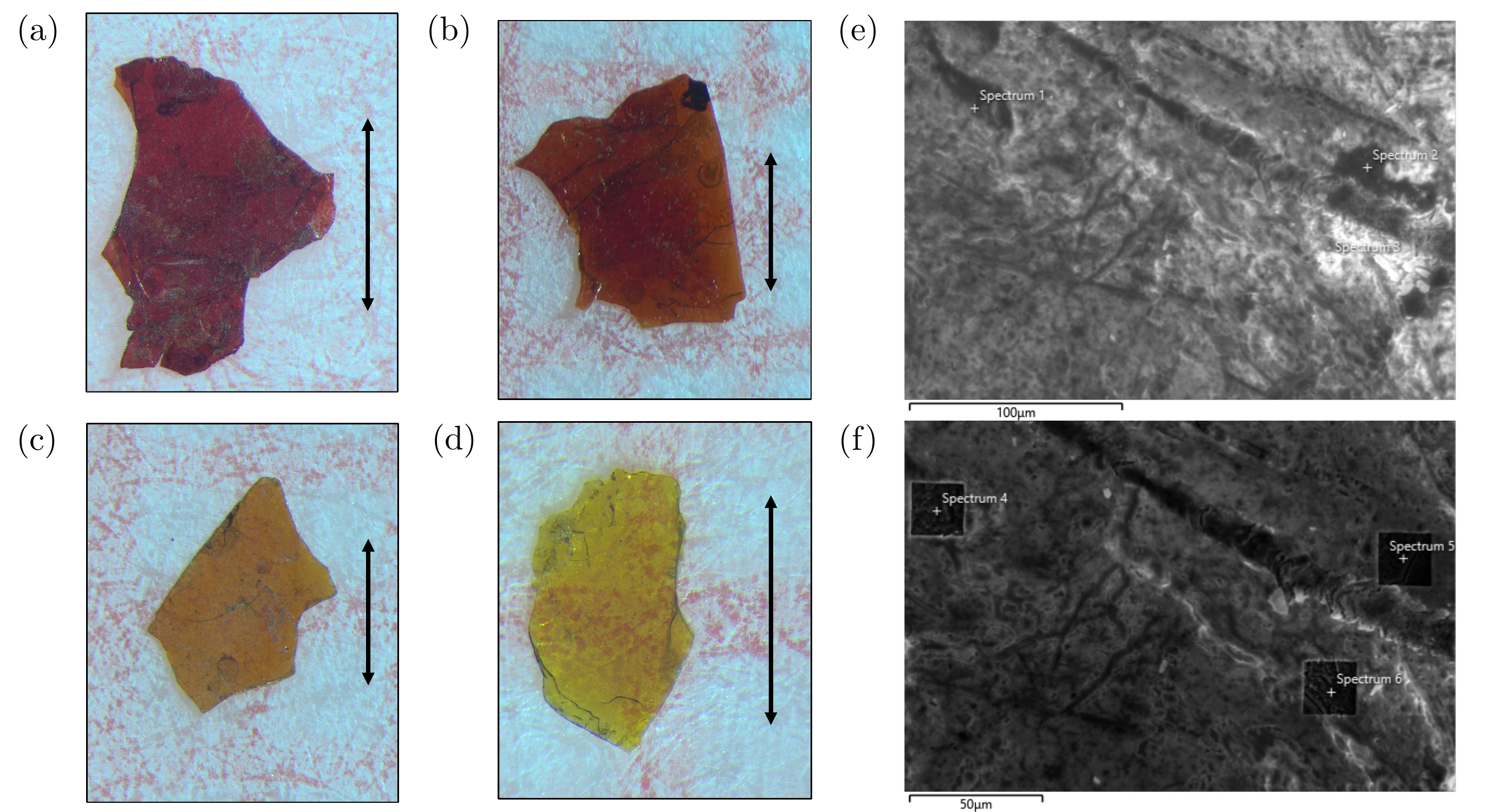}
\caption{Sample images of NaLu$_x$Yb$_{1-x}$Se$_2$, where (a), (b), (c) and (d) are for $x=0$, 0.5, 0.9 and 1 respectively. The black line is 1~mm in all cases. (e) and (f) show SEM images of a NaYbSe$_2$ sample that was left to oxidise, before and after a focused ion beam was used to remove $\sim1$~$\mu m$ of the surface in certain regions.}
\label{Samples}
\end{figure}

Single crystals of NaLu$_x$Yb$_{1-x}$Se$_2$ were grown by a NaCl flux method. Yb/Lu powder (both $>$99.9\% purity, Thermo Scientific), Se pellets ($>$99.999\%, Alfa Aesar) and NaCl crystalline powder ($>$99.0\%, VWR Chemicals) were mixed in a molar ratio of 1:2.4:28, before being sealed in a quartz tube under 200~Torr of argon. The tube was then heated to 670~$^{\circ}$C over 10 hours, held at this temperature for 10 hours, then heated to 1000~$^{\circ}$C over 15 hours and held for 7 days, before cooling to room temperature over 20 hours. The NaCl flux was then dissolved in deionised water in order to extract plate-like single crystals, up to a few mm in size (as shown in Figure~\ref{Samples}).
\\

Powder X-ray diffraction measurements were performed using a Rigaku Ultima-4 system, with Cu K$-\alpha$ radiation. The powder samples were ground from single crystals in all cases. Rietveld refinements were performed in Profex. Energy dispersive X-ray (EDX) measurements were taken using a Scios 2 DualBeam from ThermoFisher Scientific, with an accelerating voltage of 40~keV. Analysis was performed using AZtec. Raman spectra were collected at room temperature - on both exfoliated and non-exfoliated samples - using a Renishaw inVia confocal microscope equipped with a multiple-grating spectrometer and a nitrogen-cooled charge-coupled device (CCD) camera. A 100X objective with a numerical aperture of 0.85 was used to focus a 633~nm laser line onto a $\sim$1~$\mu m$ spot on the sample, and the Raman signal was collected via the same objective in the back reflection mode. The Raman spectra was calibrated against the spectrum of an internal silicon chip. To avoid laser-induced heating, the laser power was kept below 100~$\mu W$.
\\

DC magnetisation measurements below 14~T were performed using the VSM option of a Quantum Design PPMS Dynacool. Single crystal samples were attached to the sample holder (standard brass/quartz holder for fields parallel/perpendicular to the $c-$axis) using GE~7031 varnish. In all cases, the diamagnetic contribution from the sample holder was measured and subtracted from the data. Heat capacity data were also collected using a Quantum Design PPMS Dynacool, and employed a $^3$He insert in order to achieve temperatures down to 0.4~K. All measurements were performed on single crystals, with the field aligned parallel to the c-axis in all cases. \\

High-field magnetisation data was taken at the National High Magnetic Field Lab in Los Alamos. The pulsed-field magnetisation experiments used a 1.5~mm bore, 1.5~mm long, 1500-turn compensated-coil susceptometer, constructed from 50-gauge, high-purity copper wire \cite{Goddard_2008}. The susceptometer was placed within $^3$He cryostats providing temperatures down to 0.4~K. When a sample is within the coil, the signal is $V \propto (dM/dt)$, where $M$ is the magnetisation and $t$ is time. Numerical integration is used to evaluate $M$. Samples were mounted within a 1.3~mm diameter ampoule that can be moved in and out of the coil. Accurate values of $M$ are obtained by subtracting empty-coil data from those measured under identical conditions with the sample present. The magnetisation shown in Figure~\ref{Magnetisation_SM} was obtained by splicing together multiple data sets from different pulses, and the given temperature is therefore an average.


\subsection{EDX}

Table~\ref{EDX_tab} shows the composition of each growth as measured by energy dispersive X-ray (EDX) spectroscopy. A single representative sample from each growth is listed in Table~\ref{EDX_tab}, but the percentage compositions of each element are an average over multiple (at least 7) spectra taken at random points on the surface. We note that the doped compounds are slightly further from stoichiometry, and in particular, Lu is preferentially displaced in NaLu$_{0.9}$Yb$_{0.1}$Se$_2$. Excess Na is straightforwardly understood, as Lu/Yb and Na both lie in a similar CEF environment.
\\

\begingroup
\setlength{\tabcolsep}{14pt} 
\renewcommand{\arraystretch}{1.4} 
\begin{table}[h!]
\centering
\begin{tabular}{||c|c|c|c|c||} 
 \hline
$x$ & Na \% & Lu \% & Yb \% & Se \% \\ 
 \hline\hline
0 & 24.6  & - & 24.7 & 50.7 \\ 
0.5 & 25.9 &  11.8 & 11.7 & 50.6 \\ 
0.9 & 26.7  & 18.4 & 2.3 & 52.6 \\
   1 & 25.0  & 24.8 & - & 50.2 \\   
 \hline
\end{tabular}
\caption{EDX results for NaLu$_{x}$Yb$_{1-x}$Se$_2$ samples, representative of the four different growths. In all cases, the percentage composition of each element is averaged over at least 7 spectra taken for that sample.}
\label{EDX_tab}
\end{table}
\endgroup

Additionally, we performed EDX on a sample that had been left out to air for a week, and found the surface had oxidised significantly. However, upon etching approximately 1$~\mu m$ into the surface using a focused ion beam and then repeating the measurement, the oxygen had fallen to background levels. This would imply that any oxidation is constrained to a thin layer on the surface, and should not therefore affect our bulk measurements in any significant way.
\\

\subsection{PXRD}

\begin{figure}[h]
\centering
\includegraphics[width=1\textwidth]{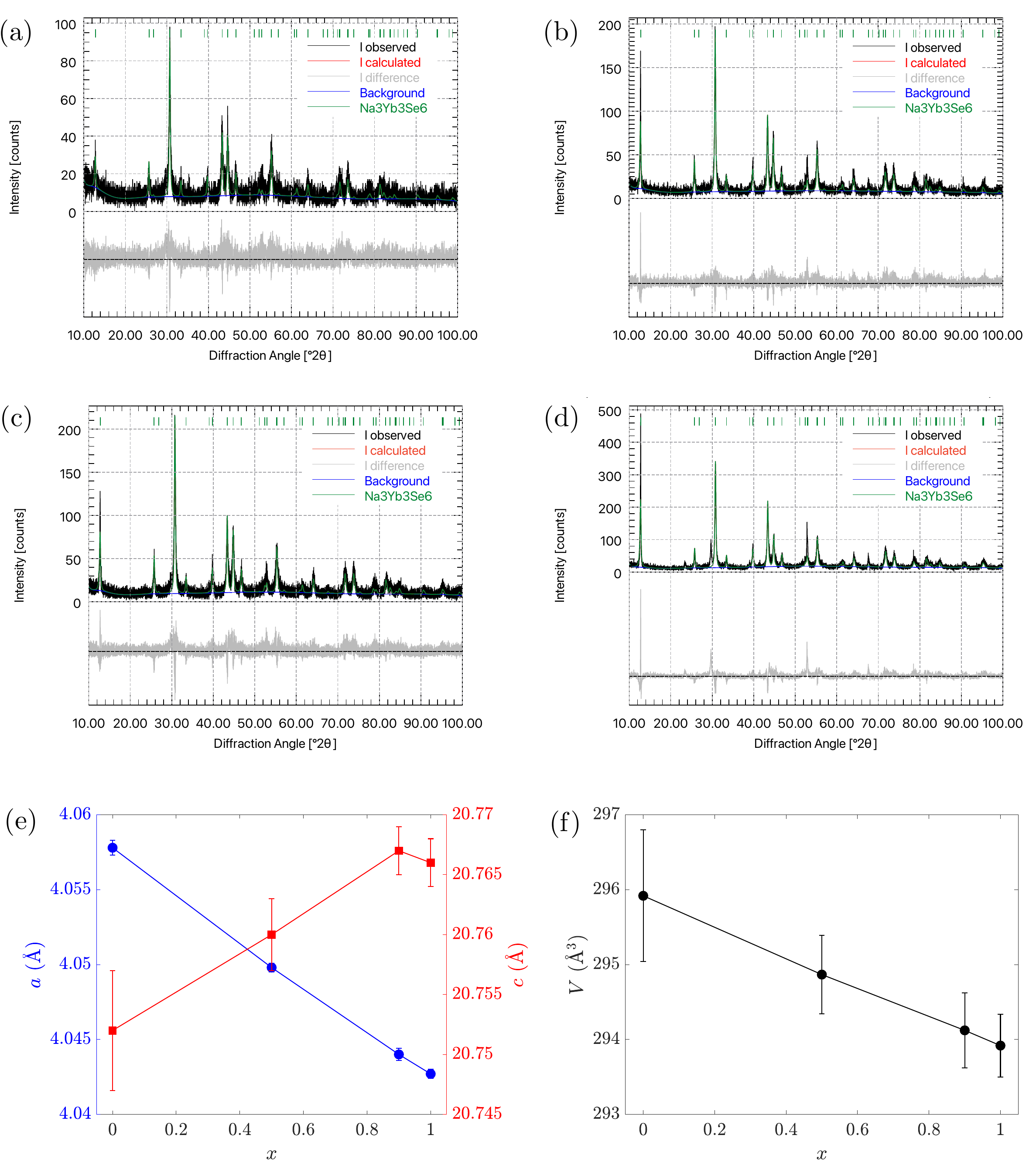}
\caption{Powder diffraction patterns and Rietveld refinements for NaLu$_x$Yb$_{1-x}$Se$_2$, where (a), (b), (c) and (d) are for $x=0$, 0.5, 0.9 and 1 respectively. (e) and (f) show the evolution of the lattice parameters and unit cell volume with varying Yb content.}
\label{PXRD}
\end{figure}

Figure~\ref{PXRD} shows the Rietveld refinements for all four compounds, alongside the evolution of the lattice parameters and unit cell volume as a function of doping.
\\

\subsection{Magnetisation}

\begin{figure}[h]
\centering
\includegraphics[width=1\textwidth]{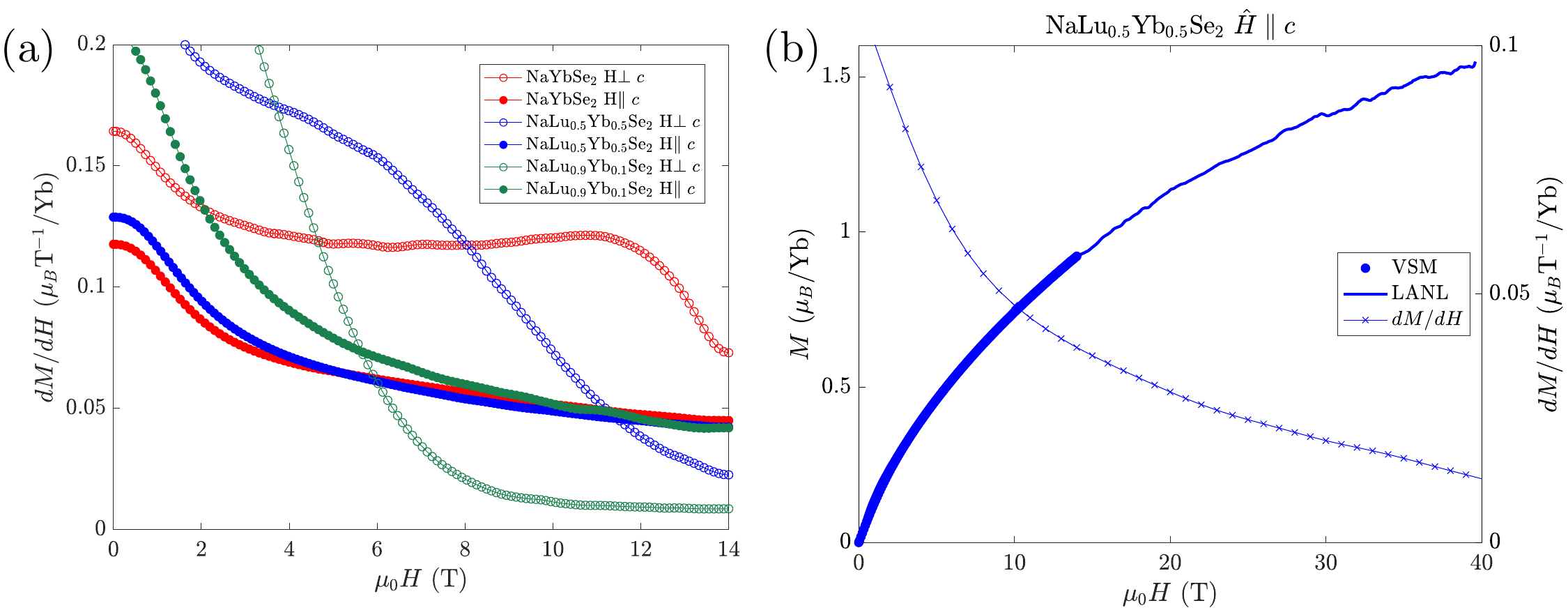}
\caption{(a) Magnetisation data (shown in the main text) differentiated with respect to field. (b) High-field magnetisation data for NaLu$_{0.5}$Yb$_{0.5}$Se$_2$ for out-of-plane fields, taken at 0.6~K. The high-field data (given by the line) has been normalised to match the low-field data (points). Also included is a smoothed differentation of the magnetisation data, taken with respect to field.}
\label{Magnetisation_SM}
\end{figure}

As discussed in the main text, the magnetisation fails to saturate in all cases bar one. This is demonstrated more clearly in Figure~\ref{Magnetisation_SM}~(a), which plots the magnetisation differentiated with respect to field. Only in one case (NaLu$_{0.9}$Yb$_{0.1}$Se$_2$ $\hat{H}\perp c$) does this saturate to a constant value, implying that the Van Vleck contribution is a small additive correction. In fact, for NaLu$_{0.5}$Yb$_{0.5}$Se$_2$ $\hat{H}\parallel c$, the magnetisation does not approach linearity even for fields as large as 40~T (Figure~\ref{Magnetisation_SM}~(b)). This is strong evidence that second order corrections are not negligible in this case, and the Van Vleck correction cannot simply be extracted as a simple linear contribution.
\\

In order to estimate these higher order corrections, we can model the low-temperature susceptibility using
\begin{equation}
\chi^{-1} = \frac{\mu_0 k_B}{N_A} \Bigg[ \frac{T}{\frac{1}{4} g^2 \mu_0^2 \mu_B^2 - 2\alpha k_B T} + \Theta_\text{CW}\Bigg],
\end{equation}
where $\alpha$ is the $H^2$ coefficient when the free energy is expanded in field. In this expression, $\chi$, $g$, $\alpha$ and $\Theta_{\text{CW}}$ will each be different for the two perpendicular field directions. The above expression is derived in \cite{Pocs2021} and is a simplification which only considers the ground state doublet. It is therefore only valid in the low-temperature limit ($T<50$~K in this case).
\\

\begin{figure}[h]
\centering
\includegraphics[width=1\textwidth]{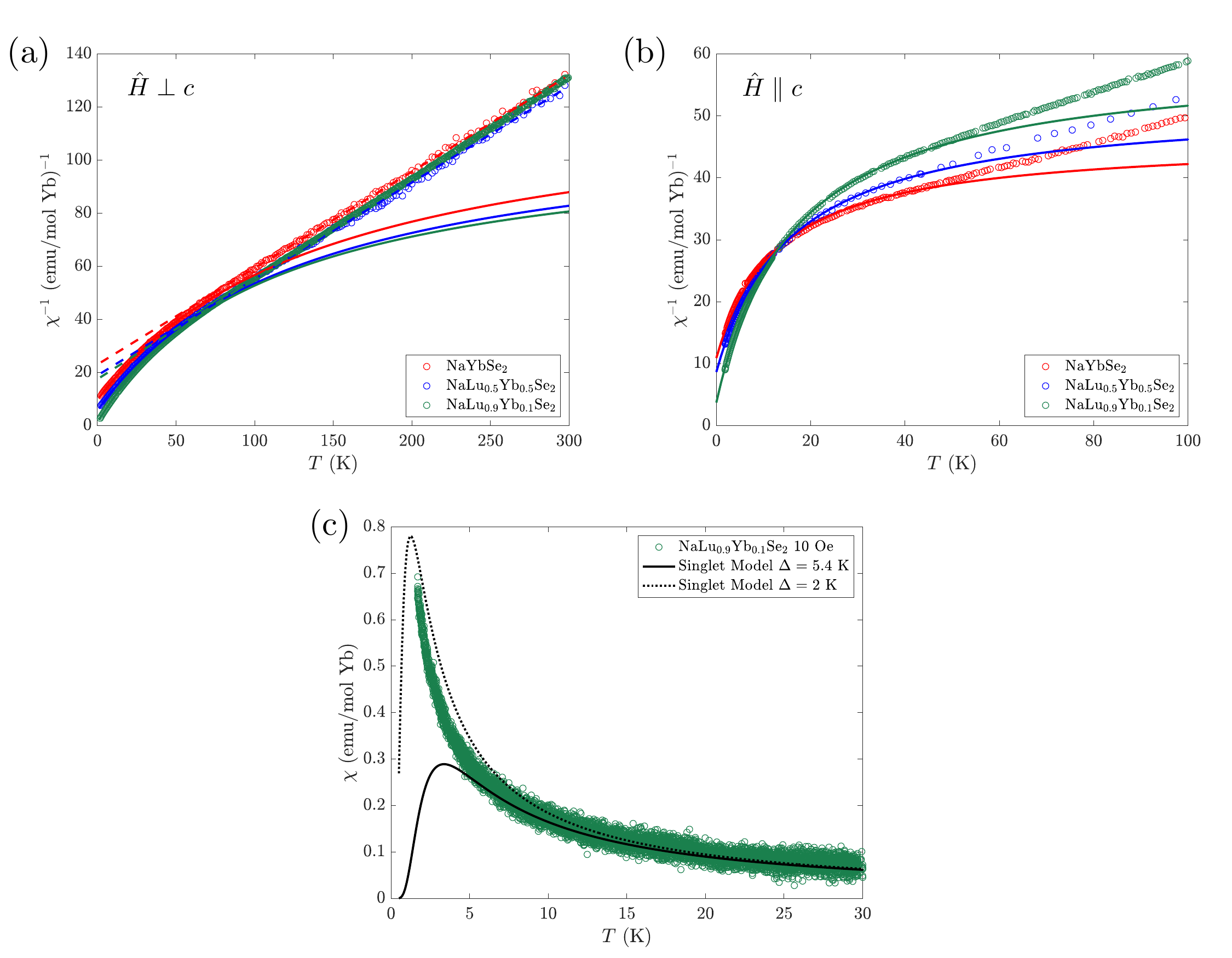}
\caption{(a) Inverse susceptibility for in-plane fields. Dashed lines show the Curie-Weiss fit for $T>100$~K, whilst solid lines show fits to a modified Curie-Weiss law which includes higher order corrections, as described in the supplementary text. (b) Inverse susceptibility for out-of-plane fields. Again, the solid lines shows the low-temperature fits including higher order corrections. (c) Magnetic susceptibility of NaLu$_{0.9}$Yb$_{0.1}$Se$_2$ for an in-plane field of 10~Oe. The curves show the susceptibility for simple fits to a dimer system with a triply degenerate excited state, with two different energy gaps.}
\label{Susceptibility_SM}
\end{figure}

Shown in Figure~\ref{Susceptibility_SM}~(a) and (b) are fits to the low-temperature susceptibility using the above expression, for in- and out-of-plane fields respectively. The fitting parameters are listed in Table~\ref{Mag_tab}. By comparing these values to the values estimated from the low-field magnetisation and specific heat in the main text (and also by previous studies \cite{Ranjith2019,ZhangModel2021,ZhangCEF2021}), this method would appear to overestimate the $g$-factor, but underestimate $\alpha$. This is also true for the related compound CsYbSe$_2$ \cite{Pocs2021}. However, it is clear that second order corrections are relevant, particularly for out-of-plane fields (although this was perhaps already evident from the low-temperature behaviour of the susceptibility). Also included in Figure~\ref{Susceptibility_SM}~(a) is a regular Curie-Weiss fit to the in-plane field, high-temperature data ($T>100$~K). As discussed in the main text, the estimated effective moment is similar to that of a free Yb$^{3+}$ ion (Table~\ref{Mag_tab}).
\\

Note, if we instead assume that the in-plane Van Vleck contribution estimated from the low-field NaLu$_{0.9}$Yb$_{0.1}$Se$_2$ magnetisation data is similar for all compounds (for $\hat{H}\perp c$), we can subtract this from the susceptibility and then perform a regular Curie-Weiss fit at $T<30$~K. This gives $g_\perp\sim3$ in all cases (similar to previous studies), and $\Theta_{\text{CW}}$ to be approximately 20\% larger (again in all cases). 
\\

Finally, in Figure~\ref{Susceptibility_SM}~(c) we plot the magnetic susceptibility of NaLu$_{0.9}$Yb$_{0.1}$Se$_2$ for an in-plane field of 10~Oe. This is in order to demonstrate that the susceptibility remains resolutely paramagnetic to the lowest measured temperatures, even in significantly smaller fields. Also included is the expected susceptibility for a spin-1/2 dimer with isotropic coupling, given by
\begin{equation}
    \chi = \frac{2N_A g^2 \mu_B^2}{k_B T} \frac{1}{3 + e^{-\Delta/T}},
\end{equation}
which is again calculated from the expansion of the free energy in $H$, assuming a single energy gap of $\Delta$ between the non-degenerate ground state and triply degenerate excited state \cite{Magnetochemistry}. We plot this curve for $\Delta=5.4$~K and $\Delta=2$~K, in order to show that there is no turn off from paramagnetic behaviour at the energy scale predicted from the zero-field specific heat measurements, but if the dimer exchange coupling was slightly smaller, the turn off would be below the lowest measured temperature.

\begingroup
\setlength{\tabcolsep}{14pt} 
\renewcommand{\arraystretch}{1.4} 
\begin{table}[h!]
\centering
\begin{tabular}{||c|c|c|c|c|c|c||} 
 \hline
$x$ & $\hat{H}$ & $\Theta_{\text{CW}}^+$ (K) & $\mu_{\text{eff}}^+$ $(\mu_B)$ & $\Theta_{\text{CW}}^-$ (K) & $g^-$ & $\alpha^-$ (10$^{-3}$ $\mu_B$T$^{-1}$) \\ 
 \hline\hline
0 & $\perp c$ & 58.9  & 4.60 & 9.7 & 3.7 & -7.6 \\ 
0.5 & $\perp c$ & 47.9 & 4.61 & 5.6 & 3.6 & -8.1 \\ 
0.9 & $\perp c$ & 41.9  & 4.51 & 1.2 & 3.3 & -8.2 \\ 
0 & $\parallel  c$ & -  & - & 11 & 2.0 & -26 \\ 
0.5 & $\parallel c$ & -  & - & 8.6 & 2.0 & -21 \\ 
0.9 & $\parallel c$ & -  & - & 3.7 & 1.8 & -16 \\ 
 \hline
\end{tabular}
\caption{List of parameters estimated from high-temperature ($^+$, $>100$~K) Curie-Weiss fits to the in-plane field inverse suscepibility, and low-temperature ($^-$, $<50$~K) modified Curie-Weiss fits to the inverse susceptibility for both field directions, for the magnetic compounds NaLu$_{x}$Yb$_{1-x}$Se$_2$, as shown in Figure~\ref{Susceptibility_SM}.}
\label{Mag_tab}
\end{table}
\endgroup

\subsection{Specific Heat}

\begin{figure}[h]
\centering
\includegraphics[width=1\textwidth]{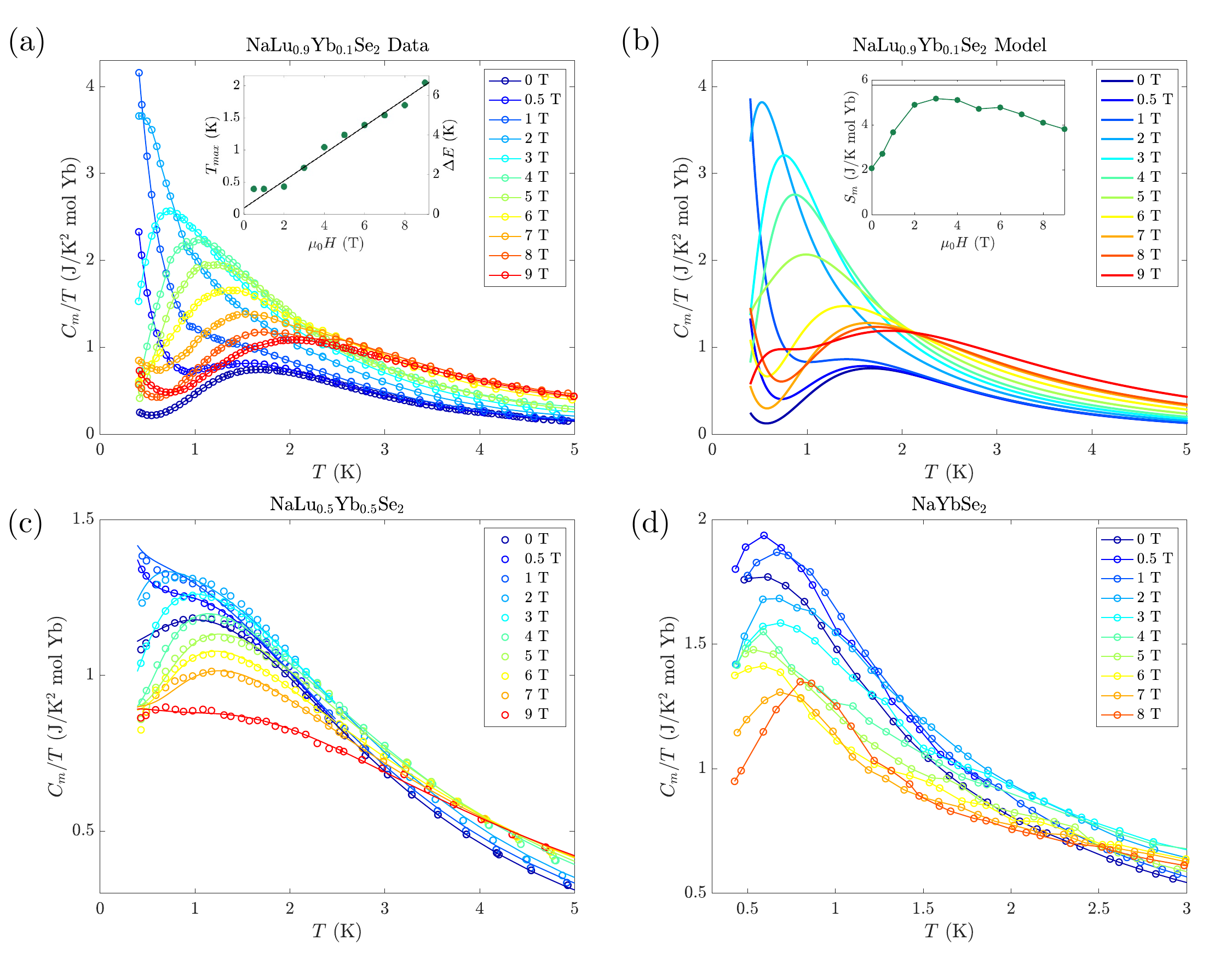}
\caption{In-field magnetic specific heat for the three magnetic compounds, plotted as a function of temperature, all for out-of-plane fields. (a) NaLu$_{0.9}$Yb$_{0.1}$Se$_2$ data, the lines join the data points as a guide to the eye. The inset shows the maximum temperature of $C_m/T$, plotted as a function of field, with the solid line a linear fit through the points above 1~T. (b) NaLu$_{0.9}$Yb$_{0.1}$Se$_2$ model, as described in the text. The inset shows the entropy released between 0.4---6~K, calculated by integrating an interpolation through the data points, plotted as a function of field. (c) NaLu$_{0.5}$Yb$_{0.5}$Se$_2$ data, with the solid lines a fitting to the Gaussian broadened two-level model plus a small Zeeman split Schottky contribution, as described in the text. (d) NaYbSe$_2$ data, with the lines simply to join the data points, as a guide to the eye again.}
\label{HC_SM}
\end{figure}

Figure~\ref{HC_SM} shows the in-field specific heat of the magnetic compounds, with an interpolation through the non-magnetic analogue NaLuSe$_2$ having been subtracted from the data. Figure~\ref{HC_SM}~(a) shows the NaLu$_{0.9}$Yb$_{0.1}$Se$_2$ data, and the inset shows the position of the maxima plotted as a function of field. From Equation~\ref{C2}, the peak position for a two-level non-degenerate system will occur at a temperature $\sim0.31\Delta$, where $\Delta$ is the energy splitting. This feature results from the Zeeman splitting of a spin-1/2 doublet, and has an energy splitting of $\Delta=g_\parallel \mu_B H_{\text{eff}}/k_B.$ From a linear fit - which allows for a zero-field splitting - we can estimate $g_\parallel=1.04$. The inset of Figure~\ref{HC_SM}~(a) shows the released entropy between 0.4---6~K (calculated from the data), which rises to almost the anticipated $R\ln{2}$ at intermediate fields - the discrepancy is simply due to the finite temperature window.
\\

In modeling the NaLu$_{0.9}$Yb$_{0.1}$Se$_2$ specific heat data (shown in Figure~\ref{HC_SM}~(b)), we assume two independent species of isolated single ions and isolated dimers. The single ions are modeled using Equation~\ref{C2}, with a Zeeman splitting $\Delta=g \mu_B H_{\text{eff}}/k_B$ as described in the main text. The Yb dimers will necessarily have four energy levels, and the specific heat is modeled using 
\begin{equation}
C(T) = R \frac{d}{dT} \bigg[\frac{ \Delta_0 e^{-\Delta_0/k_B T} + \Delta_+ e^{-\Delta_+/k_B T} + \Delta_- e^{-\Delta_-/k_B T} }{1 + e^{-\Delta_0/k_B T} + e^{-\Delta_+/k_B T} + e^{-\Delta_-/k_B T}}\bigg],
\end{equation}
where $\Delta_0=J_\perp$ is the separation between the singlet ground state and $\ket{1,0}$ triplet state, and
\begin{equation}
    \Delta_\pm = J_\parallel \mp \frac{g_\parallel \mu_B H_{\text{eff}}}{k_B} - \alpha_\pm H_{\text{eff}}^2
\end{equation}
is the separation between the singlet ground state and the excited $\ket{1,\pm1}$ doublet \cite{Pocs2021}. Note that we have included a quadratic correction term ($\alpha_\pm$), which the susceptibility data shows is significant for out-of-plane fields (Table~\ref{Mag_tab}). Note also, as this shifts the single ion energy levels in a similar fashion, the energy splitting will remain identical to the first order Zeeman splitting. Finally, there may be additional correction terms arising due to the in-plane components of $H_{\text{eff}}$, but based on the estimated size of the zero-field splitting we anticipate these to be small. 
\\

\begin{figure}[h]
\centering
\includegraphics[width=1\textwidth]{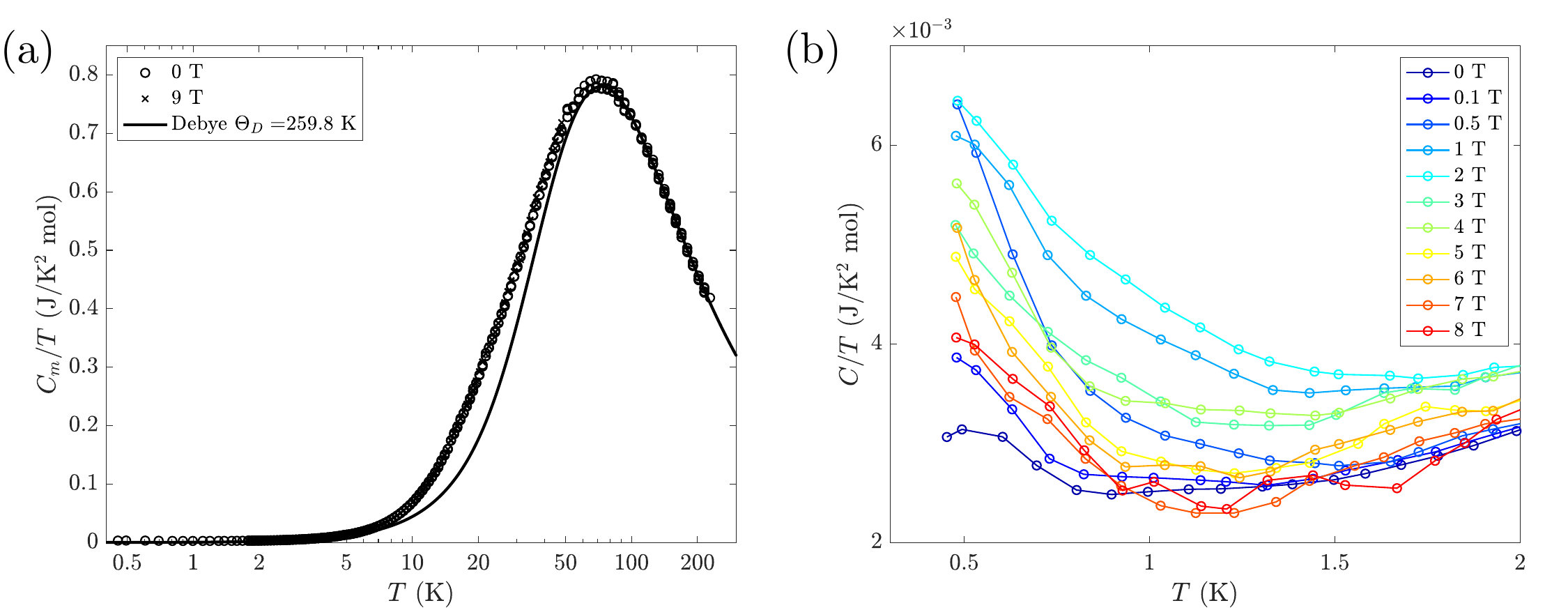}
\caption{Specific heat of NaLuSe$_2$. (a) 0 and 9 T data plotted over the full measured temperature range, showing comparison to a Debye fit with $\Theta_\text{D}=259.8$~K. (b) Low temperature, in-field specific heat data.}
\label{LuHC}
\end{figure}

Figure~\ref{LuHC} shows the specific heat of NaLuSe$_2$. The data is fit reasonably well to a Debye model with $\Theta_\text{D}=259.8$~K. Figure~\ref{LuHC}~(b) shows the low temperature, in-field data, which exhibits a small upturn. This may be attributed to the interaction of the Lu nuclear quadrupolar moment (the primary stable isotope of which has a nuclear spin $I=7/2$) with the CEF gradients, or perhaps a Schottky anomaly arising due to impurity spins. Importantly however, the upturn is negligible in comparison to the specific heat of the magnetic compounds.

\end{document}